\documentclass[prd
,preprint
,showkeys
,showpacs
,nofootinbib
,amsfonts,amsmath,amssymb
]{revtex4}
\usepackage[pdftex]{graphicx}
\usepackage{float}
\usepackage{color}
\newcommand{\lad}{\ell_\text{AdS}}

\newcommand{\ep}{\epsilon}
\newcommand{\pa}{\partial}
\newcommand{\del}{\delta}

\begin{document}
\title{Null wave front as  Ryu-Takayanagi surface }
\author{Jun Tsujimura}
\email{tsujimura.jun@a.mbox.phys.nagoya-u.ac.jp}

\author{Yasusada Nambu}
\email{nambu@gravity.phys.nagoya-u.ac.jp}
\affiliation{Department of Physics, Graduate School of Science, Nagoya
University, Chikusa, Nagoya 464-8602, Japan}

\date{March 30, 2020 ver 1}
 
\begin{abstract}
  The Ryu-Takayanagi formula provides the entanglement entropy of
  quantum field theory as an area of the minimal surface
  (Ryu-Takayangi surface) in a corresponding gravity theory.  There
  are some attempts to understand the formula as a flow rather than as
  a surface. In this paper, we propose that null rays emitted from the
  AdS boundary can be regarded as such a flow. In particular, we show
  that in spherical symmetric static spacetimes with a negative
  cosmological constant, wave fronts of null geodesics from a point on
  the AdS boundary become  extremal surfaces and therefore they can
  be regarded as the Ryu-Takayanagi surfaces.  In addition, based on
  the viewpoint of flow, we propose a wave optical formula to
  calculate the holographic entanglement entropy.
\end{abstract}

\maketitle
\tableofcontents

\section{Introduction}

It is well known that the entanglement entropy (EE) of conformal field
theory (CFT) can calculate in a corresponding gravity theory by
the Ryu-Takayanagi (RT) formula
\cite{PhysRevLett.96.181602Ryu2006,Ryu:2006ef} in AdS/CFT
correspondence \cite{GUBSER1998105,Witten199801}. In general,
although the EE of quantum field theory is not easy to calculate, the
RT formula tells that the EE $S_A$ of a region $A$ in  CFT
can calculated as the area of the minimal bulk surface $\mathcal{M}_A$ 
  homologous to $A$ (RT surface):
\begin{align}
S_A = \frac{\mathrm{Area}\left(\mathcal{M}_A\right)}{4 G_N},
\end{align}
where $G_N$ is the Newton constant of gravitation.  This relation
promotes informational theoretical analysis of AdS/CFT
correspondence. By regarding the geometry of a bulk as a tensor
network, it implements quantum error correcting code of boundary CFT
\cite{Pastawski2015} or MERA \cite{Swingle2012}, subregion subregion
correspondence which is proposed for reduced density matrix
\cite{Czech_2012}.

From this point of view, it is better to regard the RT
formula as a flow proposed by the paper \cite{Freedman:2016zud}. The
authors introduced ``bit threads'' which are equivalent concept to the
RT surface geometrically. The bit threads are defined as a bounded divergenceless
vector field $v^\mu$
\begin{align}
\nabla_\mu v^\mu=0, \quad \left|v\right| \leq C,
\end{align}
and it maximizes its flux on a boundary area $A$. The property that
the maximal total flux of $v^\mu$ through the area $A$ is equal to the
area of the RT surface is proved by the max-flow min-cut theorem
\cite{Freedman:2016zud}. The bit threads give an intuitive picture that a
vector field carrying information of the boundary propagates in the
bulk, and the bulk region stores information of the boundary.

Although the concept of bit threads is inspirational, as mentioned by
\cite{Freedman:2016zud}, the RT surface has infinitely many equivalent
bit threads. Moreover, it is non-trivial task to construct bit threads
practically and to calculate the EE of CFT by it due to its dependence
of global structure of spacetimes.  Therefore, as one of the
interpretations of the RT surface respecting the viewpoint of the
flow, we propose an interpretation that the RT surfaces are wave
fronts of null rays emitted from a point on the AdS boundary. In
particular, we prove that in spherical symmetric static spacetimes,
owing to its axisymmetry of the configuration, such wave fronts of
null rays are extremal surfaces as long as they propagates in the
vacuum region.  As the RT surface is the extremal surface
\cite{Fursaev:2006ih,Lewkowycz2013}, thus the wave front can be
considered as the RT surface. On the other hand, we can naturally
understand null rays as bit threads. In this picture, we can calculate
the EE of CFT by counting the number of such null rays. This method is
also valid for wave optical calculation using the flux of a massless
scalar field. The flux based calculation method suggests a picture
that information prepared on the boundary side spreads to the bulk as
null rays.

The structure of this paper is as follows. In Section 2, we demonstrate
the correspondence between the RT surface and the wave front of
the null rays in the BTZ spacetime.  In Section 3, we state the 
detail of our proposal and show it in spherically
symmetric static spacetimes with a negative cosmological constant.  In
Section 4, we introduce the flux formula to calculate the EE of CFT
by counting the number of null rays. Finally, Section 5 is
devoted to summary.

\section{Null wave front and RT surface}

In this section, before going to discuss the general situation, we
demonstrate that wave fronts of null rays are the RT
surface in the BTZ spacetime.

\subsection{Ryu-Takayanagi surface}

\begin{figure}[H]
\centering
\includegraphics[width=0.45\linewidth,clip]{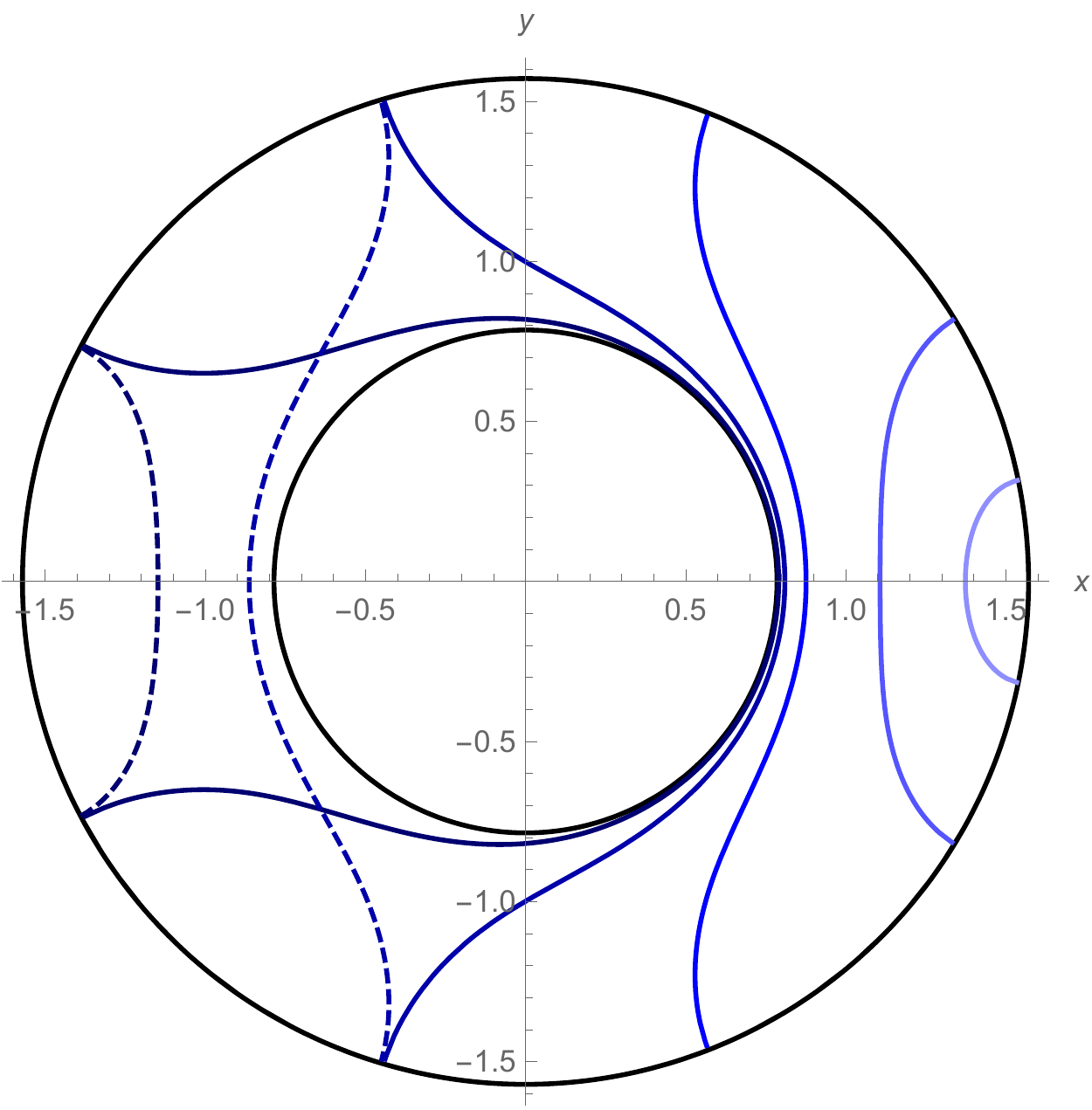}
\caption{The RT surface in the BTZ
 spacetime ($M=\lad=1$) with coordinates $(x,y)=(\rho\cos\theta,
   \rho\sin\theta)$, $\rho
  :=\lad\arctan \left(r/\lad\right)$. 
  Each blue lines is parametrized by 
  $r_{\mathrm{min}}=1.01,1.05,1.2,2.0,5.0$. For large interval $2\theta_\ell$
  on the AdS boundary, dotted lines become minimal surfaces.}
\label{RTsurfaceFig}
\end{figure}

We derive the equation of the RT surface in the BTZ spacetime
\cite{PhysRevLett.69.1849Banados1992}
\begin{align}
ds^2 = 
- \left(\frac{r^2}{\lad^2}-M\right) dt^2 +
  \left(\frac{r^2}{\lad^2}-M\right)^{-1}
  dr^2 + r^2 d\theta^2,\quad -\pi\le\theta\le\pi,
\label{eq:BTZ}
\end{align}
where $M$ is the mass of the black hole and $\lad$ is the AdS radius.
We prepare a region  (arc) $-\theta_\ell\le\theta\le\theta_\ell$ on the AdS boundary and consider a line anchored to the boundary of this region. The RT surface 
extremizes the following line area (length) on a constant time slice:
\begin{align}
\mathrm{Area}\left[r(\cdot), \frac{dr}{d\theta}(\cdot)\right] 
= \int_{-\theta_\ell}^{\theta_\ell} d\theta \sqrt{ \left(\frac{r^2}{\lad^2}-M\right)^{-1} \left( \frac{dr}{d\theta}\right)^2 + r^2 }.
\label{line element in Example}
\end{align}
 The equation of the RT surface $r=r_\text{RT}(\theta)$
is the solution of the Euler-Lagrange equation obtained by variation
of $\mathrm{Area}[r,dr/d\theta]$ with respect to $r$, and it is
\begin{align}
r_\text{RT}(\theta) = \frac{\sqrt{M}\, r_{\mathrm{min}}\,\mathrm{sech}\left(\sqrt{M}\, \theta \right)}{\sqrt{M - r_{\mathrm{min}}^2/\lad^2 \tanh^2 \left( \sqrt{M}\, \theta \right)}},
\label{eq:RT surface in Ryu-Takayanagi surface}
\end{align}
where $r_{\mathrm{min}} := r_\text{RT}(\theta=0)$ denotes the minimum of
$r$ (see Fig.~\ref{RTsurfaceFig}). Note that $\theta_\ell=\theta(r=\infty)=(1/\sqrt{M})\mathrm{arctanh}(\sqrt{M}\lad/r_\text{min})$.
The entanglement entropy of CFT on the AdS boundary for an arc $|\theta|\le\theta_\ell$  is obtained by
substituting \eqref{eq:RT surface in Ryu-Takayanagi surface} into
\eqref{line element in Example} :
\begin{align}
\mathrm{Area}\left(\mathcal{M}_\ell\right)
=2\lad\log \left( \frac{2 \lad}{\epsilon \sqrt{M}  } \sinh 
\left( \sqrt{M}\,\theta_\ell \right) \right)  + O(\epsilon),
\label{eq:finite-temp}
\end{align}
where the cutoff is introduced by
$\epsilon := \lad^2/r~ (r \rightarrow\infty)$.  Now let us
  consider CFT with inverse temperature $\beta$ on $\mathrm{S}^1$. The
  circumference of the circle is assumed to be $C$ and we prepare an
  arc $|\theta|\le\theta_\ell$ with the arc length
  $\ell=C\theta_\ell/\pi$ on it. Then it is possible to write down
  \eqref{eq:finite-temp}  using only CFT quantities. By
  dividing Eq.~\eqref{eq:finite-temp} with $4G_N$, using the
  Brown-Henneaux formula $c =3\lad/(2G_N)$ \cite{Brown1986} and
  AdS/CFT dictionary $\beta/C=1/\sqrt{M}$, we obtain the correct EE
  formula of thermal state of CFT on $\mathrm{S}^1$
  ~\cite{Calabrese:2004eu,Calabrese:2009qy} after rescaling the cutoff
  $\ep$:
\begin{equation}
  S_A=\frac{c}{3}\log\left(\frac{\beta}{\pi\ep}\sinh\left(\frac{\pi\ell}{\beta}\right)\right).
 \label{eq:EET}
\end{equation}

\subsection{Null rays and wave front}
We consider null rays emitted from a point  on the AdS boundary
and their wave fronts. Our purpose is to find out the relation between
wave fronts of null rays and the RT surface.  We consider null rays in
the spherically symmetric static spacetime
\begin{equation}
ds^2 = - f(r) dt^2 + \frac{ dr^2}{f(r)} + r^2 d\Omega_{d-1}^2,
\label{eq:metric-Ddim}
\end{equation}
where $d$ denotes spatial dimension and $d\Omega_{d-1}^2$ is the line
element of the unit sphere $\mathrm{S}^{d-1}$. We introduce
  coordinates on $\mathrm{S}^{d-1}$ as
\begin{equation}
  \begin{bmatrix}
    x_1\\x_2\\x_3\\ \vdots \\ x_{d-2}\\x_{d-1}
  \end{bmatrix}
  =
 \begin{bmatrix}  
 \cos\psi_1 \hfill\\ \sin\psi_1\cos\psi_2\hfill\\
 \sin\psi_1\sin\psi_2 \hfill \\ \vdots \\
 \sin\psi_1\sin\psi_2\cdots\sin\psi_{d-2}\cos\psi_{d-1} \\
    \sin\psi_1\sin\psi_2\cdots\sin\psi_{d-2}\sin\psi_{d-1}
    \end{bmatrix}
\end{equation}
with $0\le\psi_1,\cdots\psi_{d-2}\le\pi,~0\le\psi_{d-1}\le 2\pi$.
The line element on $\mathrm{S}^{d-1}$ is
\begin{equation}
  d\Omega_{d-1}^2=d\psi_1^2+\sin^2\psi_1(d\psi_2^2+\sin^2\psi_2(d\psi_3^2+\cdots\cdots)).
\end{equation}
As is well known, in static spherically symmetric spacetimes,
  trajectories of null geodesics stay on a spatial two dimensional
  plane. Thus we can fix coordinate values of $\psi_2,\cdots \psi_{d-1}$ and
  assume the following (2+1)-dimensional metric to investigate wave
  fronts of null rays emitted from a point:
\begin{equation}
  ds^2=-f(r)dt^2+\frac{dr^2}{f(r)}+r^2d\theta^2,\quad \theta:=\psi_1.
  \label{eq:metric3}
\end{equation}
\begin{figure}[H]
  \centering
  \includegraphics[width=0.6\linewidth]{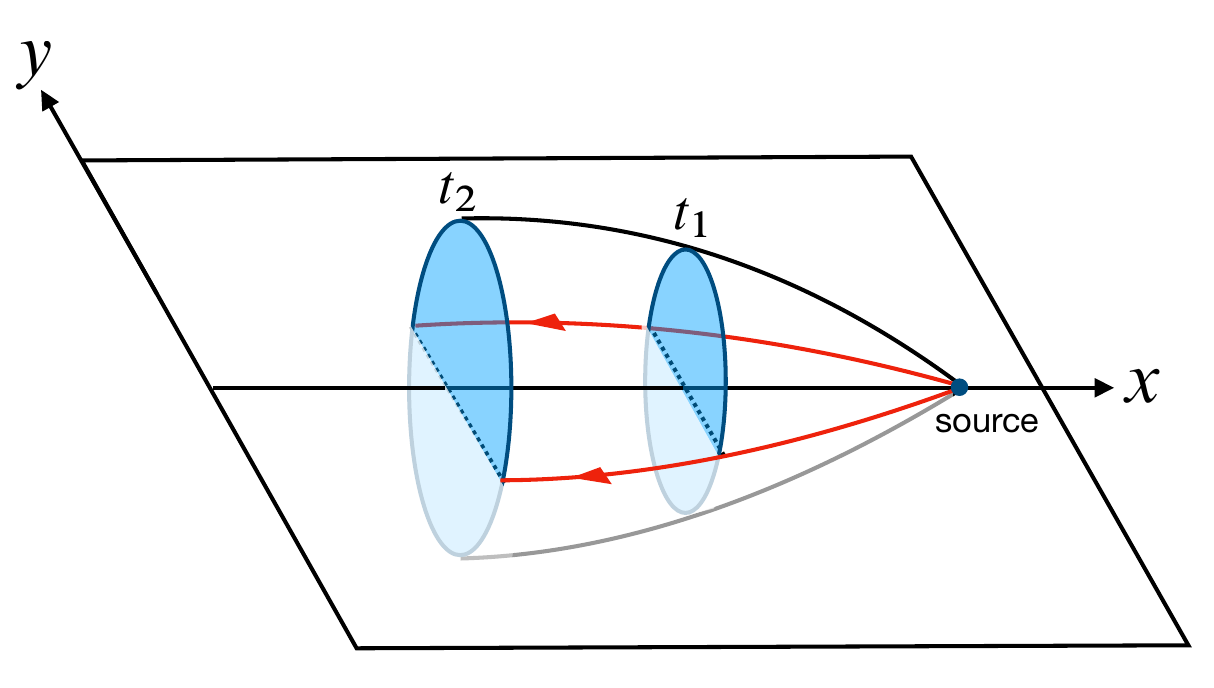}
  \caption{Axisymmetric configuration of congruence of null
    rays. Coordinates on two dimensional plane are introduced by
    $(x,y)=(r\cos\theta, r\sin\theta)$. Wave fronts at fixed $t$ are
    symmetric with respect to rotation about $x$-axis. Projection of null
    rays on the $(x,y)$-plane are shown as red lines.}
\end{figure}
\noindent
In this metric with coordinates $x^\mu=(t,r,\theta)$, a wave front of
null rays emitted from a point source is defined as a $t=$ constant
section of null congruences, which forms a $(d-1)$-dimensional
surface. Due to the axial symmetry of the configuration, a wave front
of null rays is represented as a curve in $(r,\theta)$ space in the
present situation.  The tangent vector of a null ray is
\begin{equation}
  k^\mu=(k^t, k^r, k^\theta)=\left(\frac{dt}{d\lambda},\frac{dr}{d\lambda},\frac{d\theta}{d\lambda}
\right),\quad
  -f\left(\frac{dt}{d\lambda}\right)^2+\frac{1}{f}\left(\frac{dr}{d\lambda}\right)^2+r^2\left(\frac{d\theta}{d\lambda}\right)^2=0, \label{eq:geod1}
\end{equation}
where $\lambda$ is the affine parameter.  This spacetime has two
Killing vectors related to translation of $t$ and $\theta$ directions
and there exist two conserved charges
$ \omega := f(r)\dfrac{dt}{d\lambda},\ p_\theta := r^2
\dfrac{d\theta}{d\lambda}$. Combining with Eq.~\eqref{eq:geod1}, we 
obtain a trajectory of a null ray as
\begin{align}
\theta(r) &=  \theta_0 \pm  \int_{r_0}^r \frac{b}{r'^2}\left(
            1 - f(r')
       \frac{b^2}{r'^2}\right) ^{-\frac{1}{2}} dr',
\label{eq:geo-th} \\
t(r) &=  t_0 \pm  \int_{r_0}^r \frac{1 }{f(r')}
\left(1 - f(r') \frac{b^2}{r'^2}\right) ^{-\frac{1}{2}} dr',
\label{eq:geo-t} \\
\lambda(r)&= \lambda_0 \pm \frac{1}{\omega}\int_{r_0}^r \left(1
 - f(r') \frac{b^2}{r'^2} \right)^{-\frac{1}{2}} dr',
\label{eq:geo-lam}
\end{align}
where
$(t_0,\theta_0,\lambda_0 )=
(t(r_0),\theta(r_0),\lambda(r_0) )$ and the impact parameter
$b:=p_\theta/\omega$ is introduced. The sign $\pm$ in front of
  the integral corresponds to the sign of $dr/d\lambda$.
 
\begin{figure}[H]    
\centering
\includegraphics[width=0.4\linewidth]{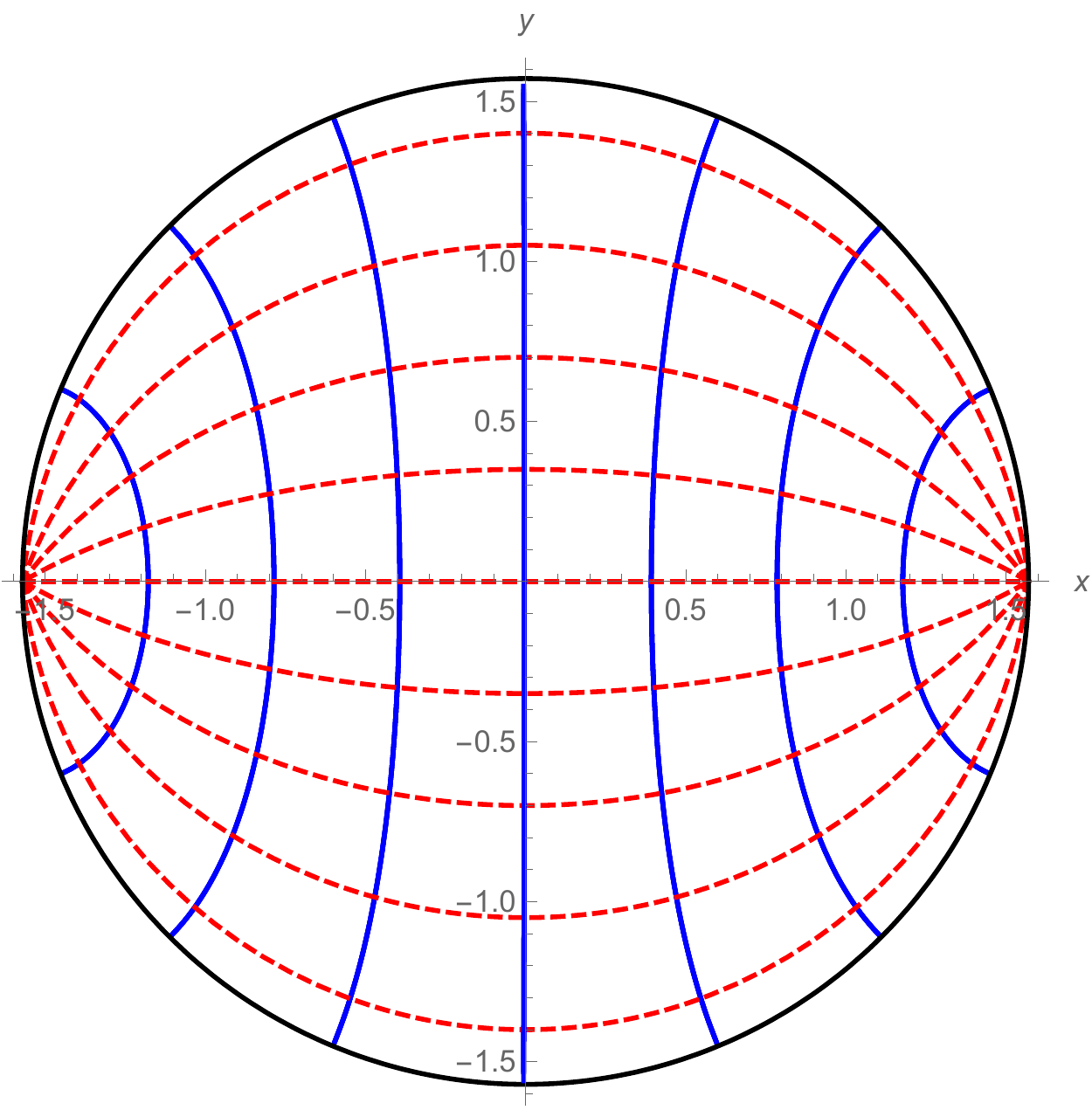}
\caption{Null rays (dotted red lines) and wave fronts (blue lines)
in the pure AdS$_{2+1}$ spacetime $(M=-1, \lad=1)$.}
\label{fig:nullAdS}
\end{figure}

\begin{figure}[H]
\centering
\includegraphics[width=0.4\linewidth]{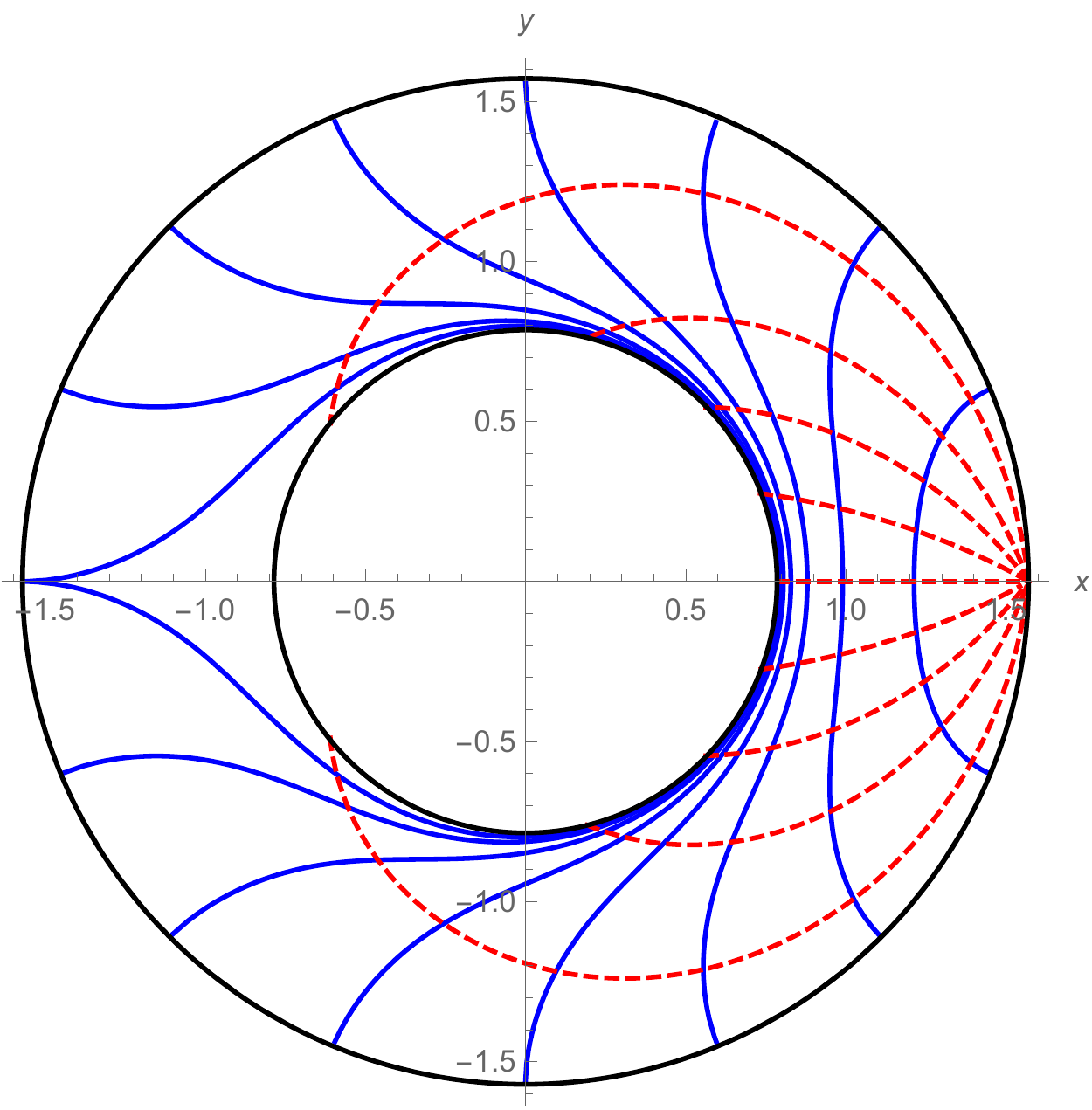}
\hspace{1cm}
\includegraphics[width=0.4\linewidth]{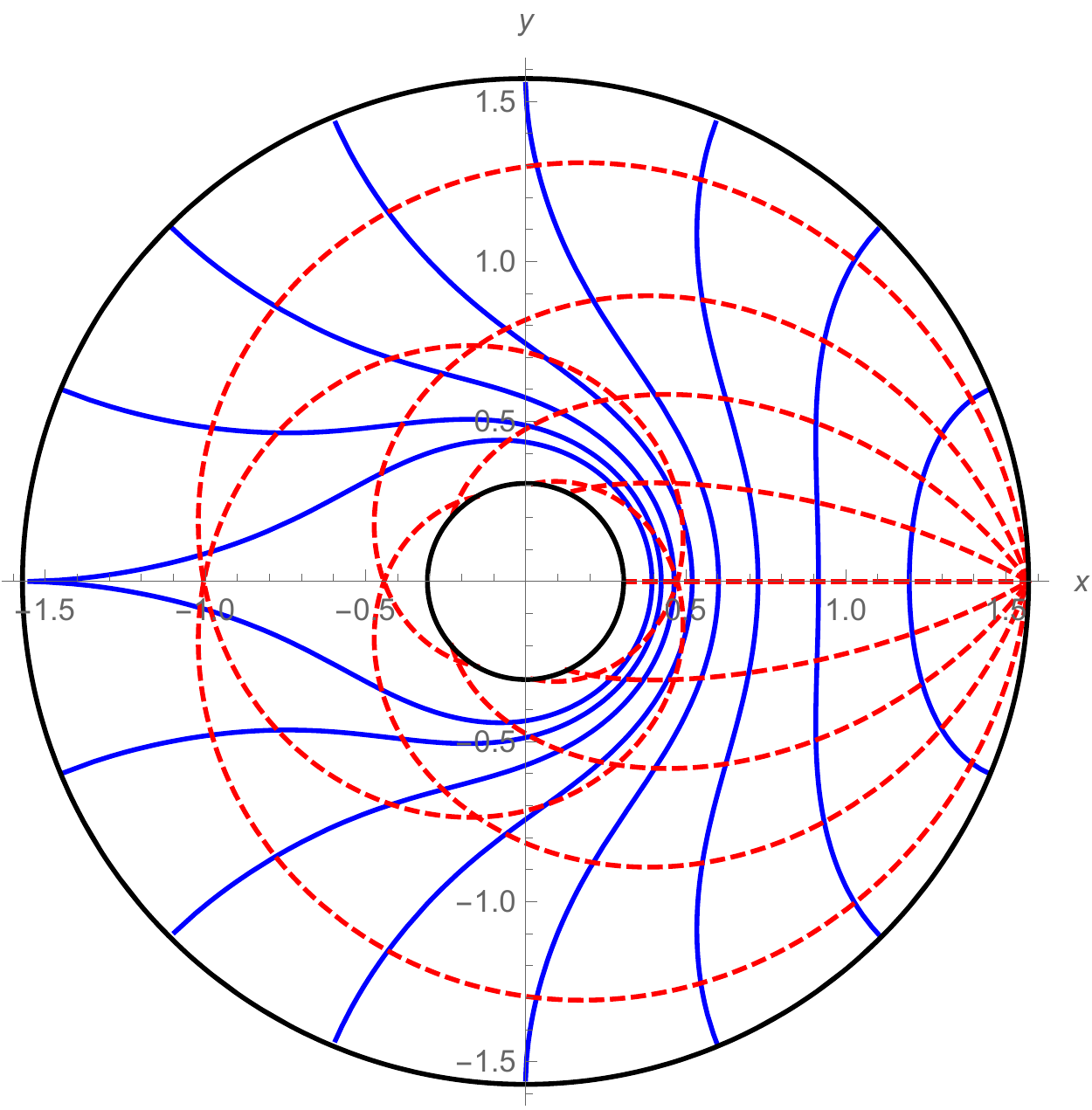}
\caption{Null rays (dotted red lines) emitted from
  $(r,\theta)=(\infty,0)$ and their wave fronts (blue lines) in the
  BTZ spacetime (left panel: $M=\lad=1$, right panel: $M=0.1,\lad=1$).}
\label{fig:nullBTZ}
\end{figure}

For the $(2+1)$-dimensional BTZ spacetime \eqref{eq:BTZ}, we can
demonstrate explicitly that wave fronts of null rays are the RT
surfaces.  We obtain equations of null geodesic from \eqref{eq:geo-th}
and \eqref{eq:geo-t} with $(t_0,r_0,\theta_0 )= (0,\infty,0)$: 
\begin{align}
\theta(r) &= \frac{1}{\sqrt{M}} \log \left[  \frac{\sqrt{r^2 - b^2\left(r^2/\lad^2 - M\right)} + b\sqrt{M}}{r\sqrt{1-b^2/\lad^2}} \right],  \label{eq:th}\\
t(\theta) &= \frac{\lad}{\sqrt{M}} \mathrm{arctanh}\left( 
\frac{\lad}{b} \tanh \left( \sqrt{M}\, \theta \right)\right). \label{eq:t}
\end{align}
It is easy to derive a trajectory of a null ray
$r=r_\text{NG}(\theta,b)$ with an impact parameter $b$ from
\eqref{eq:th}. On the other hand, the equation of a wave front
$r=r_\text{WF}(\theta,t)$ at a fixed $t$ is derived by eliminating the
parameter $b$ from \eqref{eq:th} and \eqref{eq:t}. After all,
\begin{align}
r_\text{NG}(\theta,b)&= \frac{\sqrt{M}\, b}{\sqrt{1-b^2/\lad^2}}\, 
\mathrm{csch}\left( \sqrt{M}\, \theta\right),
\label{Null geodesic Planar BTZ in Examples} \\
r_\text{WF}(\theta,t)&=  \frac{\sqrt{M}\lad\, \mathrm{coth} \left(\sqrt{M}\,t/\lad \right)\mathrm{sech} \left( \sqrt{M}\, \theta \right)}{\sqrt{ 1-\mathrm{coth}^2\left(\sqrt{M}\,t/\lad\right)\mathrm{tanh}^2 \left( \sqrt{M}\, \theta \right)} }.
\label{Wave front Planar BTZ in Examples}
\end{align}
\noindent
For the special case $M=-1$, the spacetime reduces to the pure
AdS.  Figure \ref{fig:nullAdS} and Figure
\ref{fig:nullBTZ} show null rays and their wave fronts in
the pure AdS spacetime and the BTZ black hole spacetime, respectively.
\noindent
Note that Eq.~\eqref{Wave front Planar BTZ in Examples} of the wave
front is the same as Eq.~\eqref{eq:RT surface in Ryu-Takayanagi
  surface} of the RT surface by identifying
$r_{\mathrm{min}} = \sqrt{M}\lad\coth \left( \sqrt{M} t/\lad \right)$,
which represents the elapsed time of a
null ray traveling from $r = \infty $ to
$r = r_{\mathrm {min}} $.  Indeed this quantity is obtained by
taking $b=0$ in the equation of the null ray \eqref {eq:geo-t}:
\begin{align}
  t = \int_\infty^{r_{\mathrm{min}}} \frac{dr}{r^2/\lad^2 - M} 
  = \frac{\lad}{\sqrt{M}} \!\  \mathrm{arctanh}\left(\frac{\sqrt{M}\lad}{r_{\mathrm{min}}} \right).
\end{align}
Therefore, we have confirmed that wave fronts of null rays emitted
from the AdS boundary coincide with the RT surfaces in the BTZ
spacetime. For sufficient elapse of time after emission of null rays,
self-intersection of the wave front occurs and identification of the
wave front as the RT surface becomes ambiguous.


\section{Null wave front and extremal surface}

In this section, based on the observation in the previous section for
the BTZ spacetime, we show the following proposition for spherically
symmetric static spacetimes with a cosmological constant (no matter
fields).

\vspace{0.2cm}
\noindent
\textbf{Proposition} ~Wave fronts of null rays emitted from a point
 are  extremal surfaces when the affine parameter of null rays
goes to infinity.

\vspace{0.2cm}
\noindent
\textbf{Corollary} ~For spacetimes with a negative cosmological
constant, wave fronts of null rays emitted from a point  on the
AdS boundary are  extremal surfaces.

\vspace{0.2cm}
\noindent
We adopt the metric \eqref{eq:metric3} with coordinates
$x^\mu=(t,r,\theta,\cdots)$.  Let $\xi^\mu = (\pa_t)^\mu$ be the
time-like Killing vector, $k^\mu=dx^\mu/d\lambda$ be the tangent
vector of null geodesics. We introduce the projection tensor
$P^{\mu\nu}=g^{\mu\nu} - \xi^{-2} \xi^\mu \xi^\nu=\mathrm{diag}(0,
f,1/r^2,\cdots)$
onto a constant time slice. We denote the tangent vector of null
geodesics projected onto the hypersurface as
$\tilde{k}^i= P^{i}{}_{j}\, k^j=(k^r, k^\theta,0,\cdots)$.  The
conserved quantity associated with the Killing vector is
$\omega = - \xi^\mu k_\mu=fk^t$ and the norm of the spatial vector
$\tilde{k}^i$ is $\tilde{k}^i \tilde{k}_i = \omega^2/f$.

We prove the proposition by using the fact that the extremal surface
is a surface with zero mean curvature. The mean curvature $H$ of a
wave front of null rays on a constant time slice is defined by
\begin{equation}
 H:=D_i \tilde n^i,\quad \tilde n^i=\frac{\tilde k^i}{\tilde
   k}=\frac{f^{1/2}}{\omega}(k^r,k^\theta,0,\cdots),    
\end{equation}
where $\tilde n^i$ is the unit normal vector of the wave front
and $D_i = P_i{}^j\nabla_j=(\nabla_r,\nabla_\theta,\cdots)$
is the covariant derivative on a constant time slice. Then,
\begin{align}
  H&=D_i\left(\frac{f^{1/2}}{\omega}k^i\right)=\frac{1}{f^{-1/2}\sqrt{h}}\pa_i\left(\frac{f^{1/2}}{\omega}f^{-1/2}\sqrt{h}\,k^i\right)\notag\\
  &=\frac{f^{1/2}}{\omega\sqrt{h}}\left(\pa_r(\sqrt{h}\,k^r)+\pa_\theta(\sqrt{h}\,k^\theta)\right), 
\end{align}
where $\sqrt{h}=r^{d-1}$ comes from determinant of the metric on $\mathrm{S}^{d-1}$.
On the other hand, the expansion of a null congruence is
\begin{align}
  \Theta&=\nabla_\mu k^\mu=\frac{1}{\sqrt{h}}\pa_\mu(\sqrt{h}\,k^\mu) \notag\\
  &=\frac{1}{\sqrt{h}}\left(\pa_r(\sqrt{h}\,k^r)+\pa_\theta(\sqrt{h}\,k^\theta)\right). 
\end{align}
Therefore $H=(f^{1/2}/\omega)\Theta$ and the mean curvature $H$ of a
wave front is proportional to the expansion of the null geodesic
congruence. The expansion $\Theta$ along a null geodesic obeys
  the Raychaudhuri equation
\begin{equation}
  \frac{d \Theta}{d\lambda} = -\frac{\Theta^2}{d-1}-R_{\mu\nu}k^\mu k^\nu.
  \label{eq:Raych}
\end{equation}
In the present case, as the congruence of null geodesics has axial
symmetry, the shear and the rotation of the congruence do not appear
in this equation. For vacuum spacetimes with a cosmological
constant, the term with the Ricci curvature disappears.  Then the
solution of Eq.~\eqref{eq:Raych} is
$\Theta(\lambda) = (d-1)/(\lambda-\lambda_0)$ where $\lambda_0$ is the
affine parameter at the source. Thus the expansion goes to zero as the
affine parameter goes to infinity, and the mean curvature of the wave
front is zero and is the extremal surface. Therefore the
proposition is proved.  As an example of this proposition, let us
consider a wave front in the Minkowski spacetime. A spherical wave front
emitted from a point source placed at the spatial infinity becomes
plane wave,  which is zero mean curvature surface in
the Minkowski spacetime. However, in this case, the coordinate time
\eqref{eq:geo-t} becomes infinite when a wave front of null rays
arrives at an observer.

Asymptotically AdS spacetimes are peculiar because they have the
timelike boundary. We consider the pure AdS spacetime of which metric
function is given by $f(r) = 1+r^2/\lad^2$. As
  $f\approx r^2/\lad^2$ in the vicinity of the AdS boundary, the affine
  parameter of null rays \eqref{eq:geo-lam} from the AdS boundary
  $r_0=\lad^2/\ep, \ep\rightarrow 0$ diverges as
\begin{align}
  \lambda(r)\approx\frac{1}{\omega}\int_r^{\lad^2/\ep}\left(1-\frac{b^2}{\lad^2}\right)^{-1/2}dr=\frac{\lad^2/\ep-r}{\omega\sqrt{1-b^2/\lad^2}} \to \infty.
\end{align}
On the other hand, the coordinate time \eqref{eq:geo-t}
 converges as
\begin{align}
t(r) \approx \int_r^{\lad^2/\ep}dr\frac{\lad^2}{r^2}\left(1-\frac{b^2}{\lad^2}\right)^{-1/2} =\frac{\lad^2}{\sqrt{1-b^2/\lad^2}}\,\frac{1}{r}.
\end{align}
This property also holds for general asymptotically AdS spacetimes
because they have the same metric in the vicinity of the AdS boundary
as the pure AdS spacetime. After all, we conclude that for static
spherical symmetric asymptotically AdS spacetimes, wave fronts of a
null geodesic congruence emitted from a point source on the AdS
boundary are  extremal surfaces.

\section{Flux formula}
\begin{figure}[H]
\centering
\includegraphics[width=0.4\linewidth,clip]{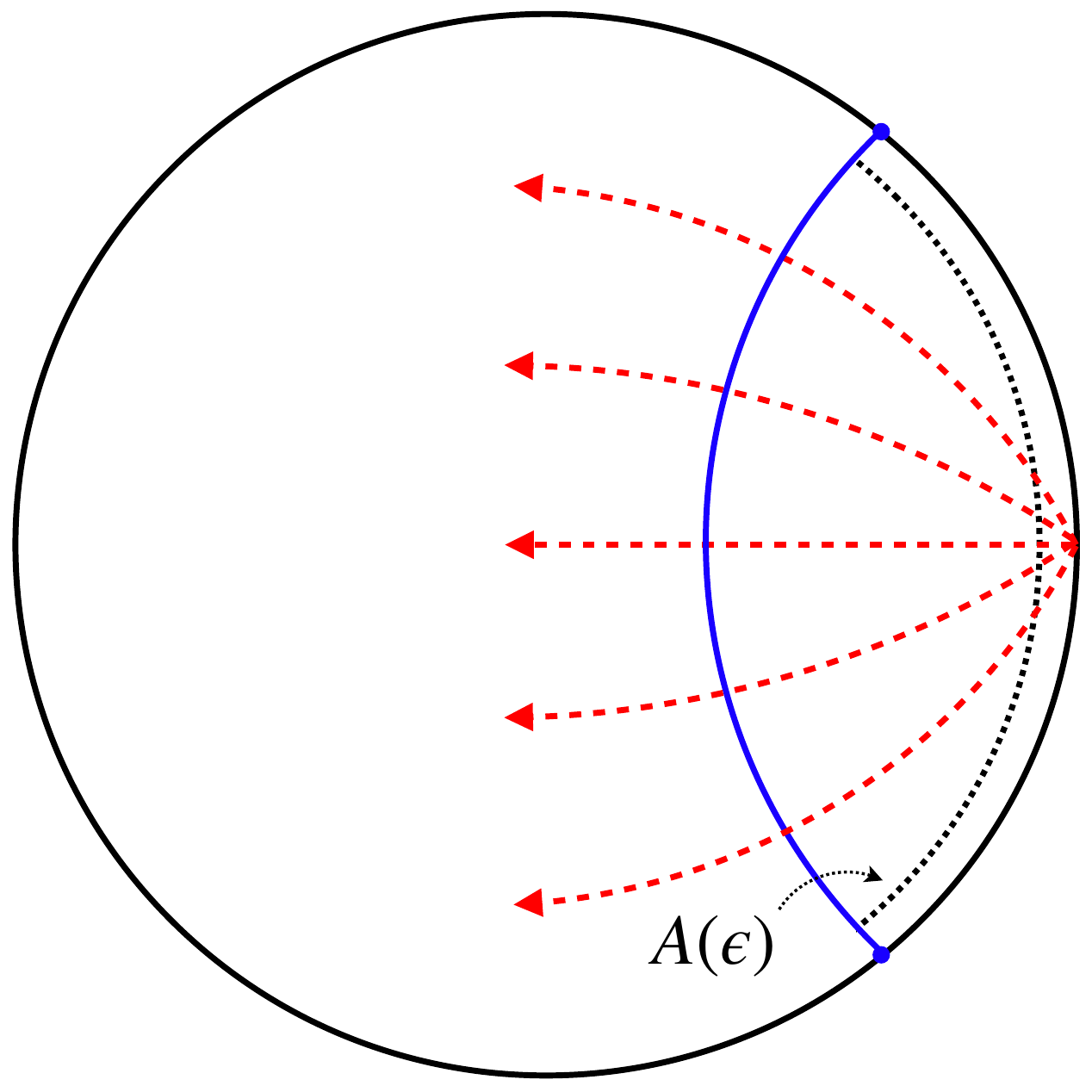}
\caption{Null rays (red dotted lines) emitted from a point on the AdS boundary
  pass through the screen $A(\ep)$ placed at $r = \lad^2/\epsilon $
  (dotted line). As the null rays are orthogonal to the wave front (blue line),
  the number of null rays is proportional to the area of the RT
  surface.}
\label{fig:screen-setup}
\end{figure}

Based on the idea presented in the previous section, we can understand
null rays as a natural flow characterizing the EE of the dual
CFT. Hence a congruence of null rays is one of the bit threads
described in Section I.  This makes us conceive a picture that null
rays propagate in the bulk with information of the AdS boundary. This
picture suggests that the EE can be calculable by counting the number
of null rays. In this section, we reformulate the RT formula in terms
of the wave optics. Concepts of wave fronts and the flux of null rays
are naturally derived as the eikonal limit of wave optics. As an
application of wave optics to black hole spacetimes, papers
\cite{Kanai2013,Nambu2016,Hashimoto2020} investigate image formation
of the photon sphere of black holes. In this paper, we focus on the
structure of wave fronts of a massless scalar field. For the
monochromatic massless scalar field with time dependence
$\propto e^{-i\omega t}$, we present wave patterns in
Fig.~\ref{fig:wave-ads} and Fig.~\ref{fig:wave-BTZ} (see detail in
Appendix). They show wave fronts from a point wave source on the AdS
bounary (see Fig.~\ref{fig:nullAdS} and Fig.~\ref{fig:nullBTZ} for
corresponding wave fronts in the geometrical optics).
\begin{figure}[H]
  \centering
  \includegraphics[width=0.4\linewidth,clip]{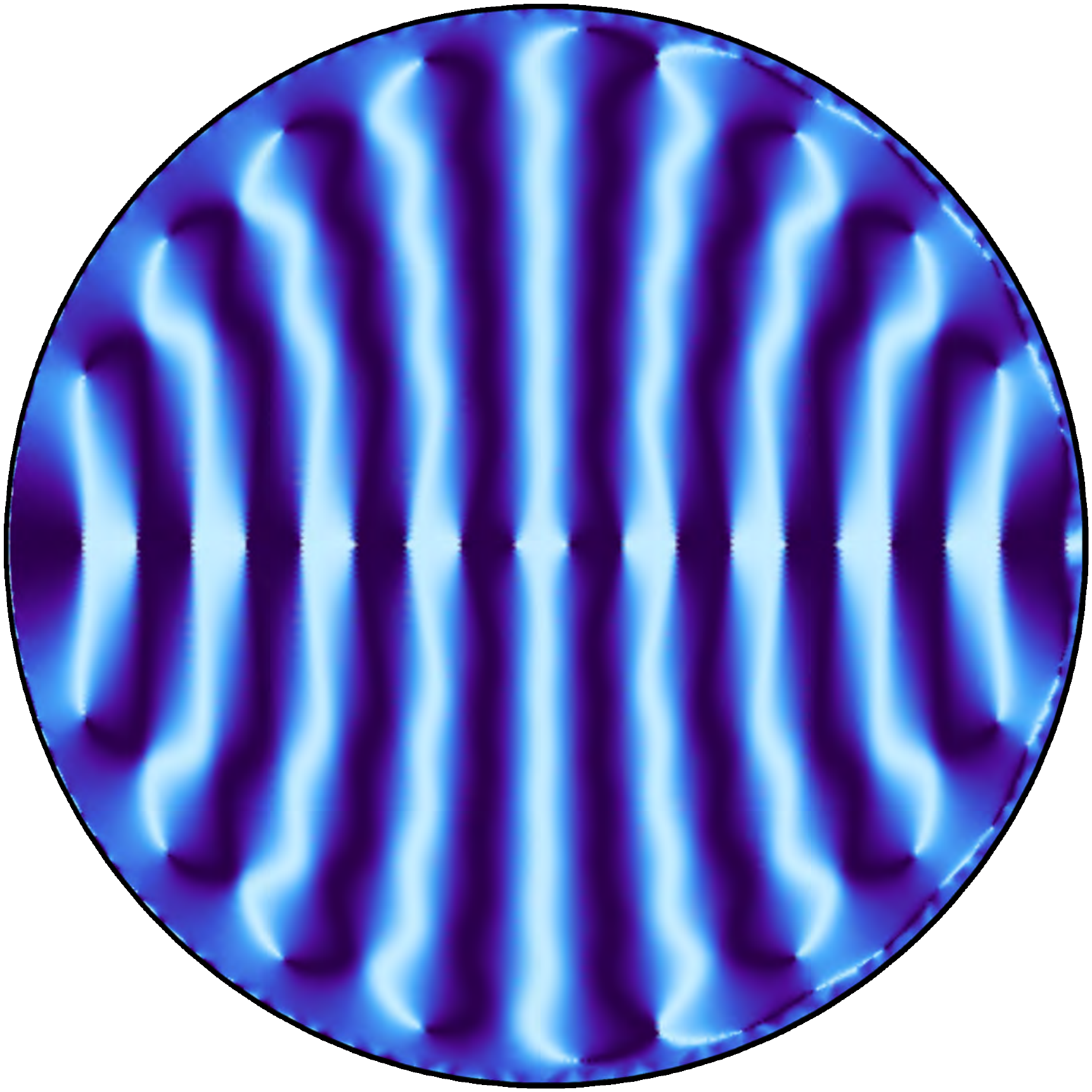}
  \caption{Wave pattern of the monochromatic massless scalar field
    with $\omega=20$  in the AdS
    spacetime. Real part of $\phi/|\phi|$ is shown.}
  \label{fig:wave-ads}
\end{figure} 
\begin{figure}[H]
  \centering
  \includegraphics[width=0.4\linewidth,clip]{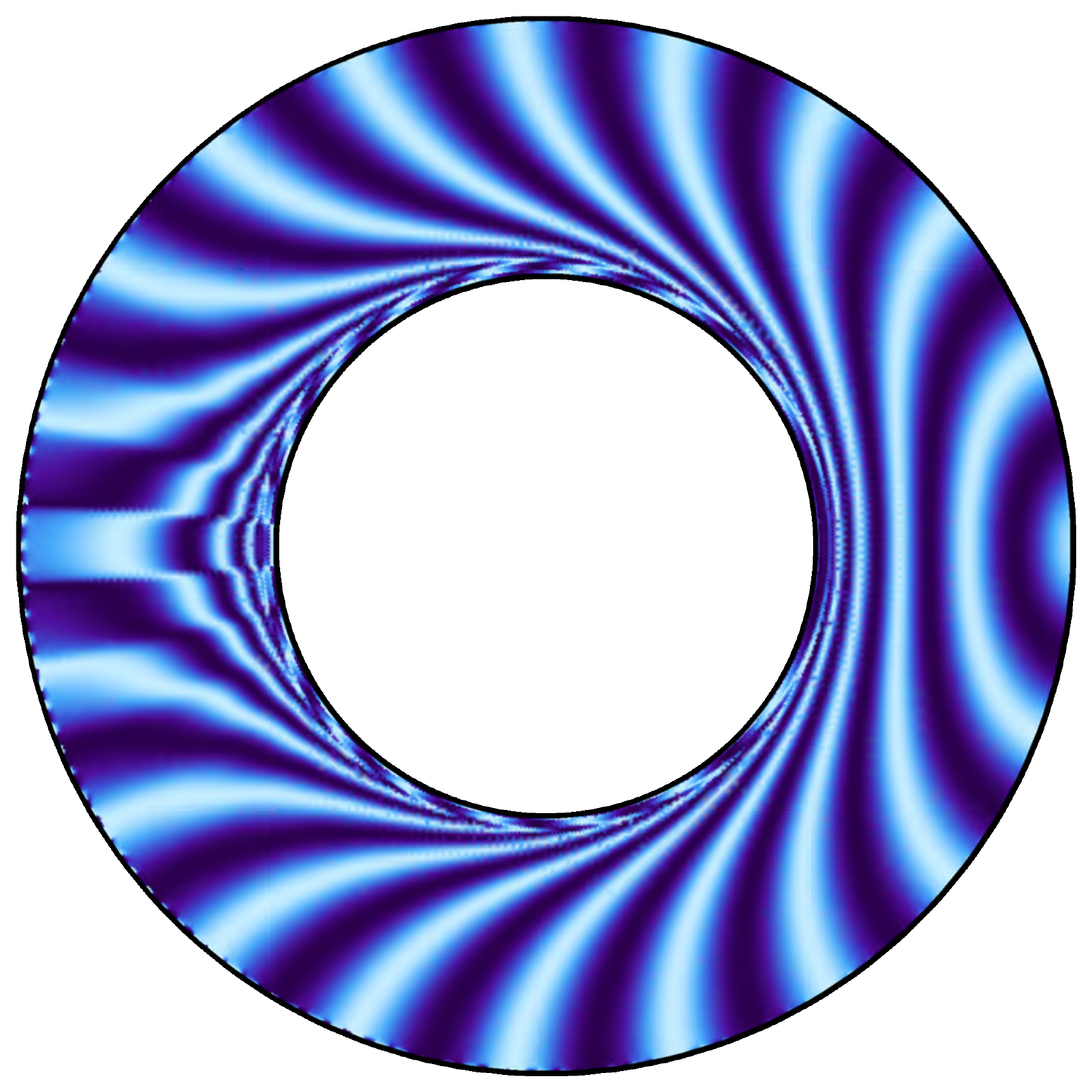}
  \hspace{1cm}
  \includegraphics[width=0.4\linewidth,clip]{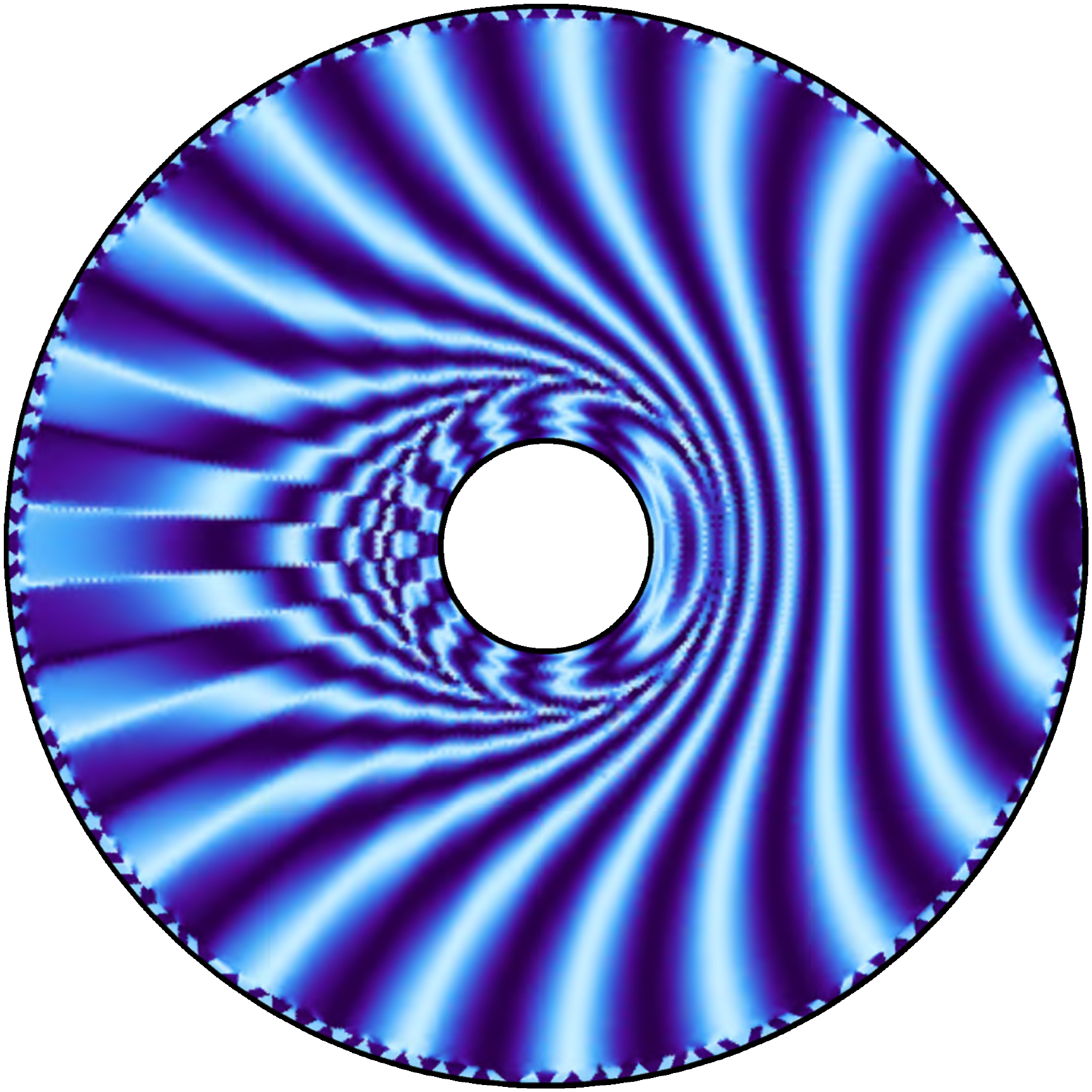}
  \caption{Wave pattern of the massless scalar field with $\omega=20$ (Re
    $\phi/|\phi|$) in the BTZ spacetime with $M=1$ (left panel) and
    $M=0.1$ (right panel).}
  \label{fig:wave-BTZ}
\end{figure}
\noindent
For the massless scalar field $\phi(x^\mu)$ obeying the Klein-Gordon
equation
$\square\phi=(\sqrt{-g})^{-1}\pa_\mu(\sqrt{-g}g^{\mu\nu}\pa_\nu\phi)=0$,
the WKB form of the wave function is
\begin{equation}
  \phi(x^\mu)=a(x^\mu)\exp\left[iS(x^\mu)\right],
\end{equation}
where $a$ and $S$ are real functions. In the eikonal limit, they obey
\begin{align}
  &g^{\mu\nu}\nabla_\mu S \nabla_\nu S=0, \label{eq:HJ} \\
  &\nabla_\mu(a^2\nabla^\mu S)=0. \label{eq:KGJ}
\end{align}
The equation \eqref{eq:HJ} is the Hamilton-Jacobi (HJ) equation and
Eq.~\eqref{eq:KGJ} represents conservation of the Klein-Gordon current
$J^\mu=(1/2i)(\phi^*\pa^\mu\phi-\phi\pa^\mu\phi^*)$. In terms of the wave vector
$k_\mu=\pa_\mu S$, which defines the tangent  of null rays,
\begin{equation}
  g_{\mu\nu}k^\mu k^\nu=0,\quad\nabla_\mu(a^2k^\mu)=0.
\end{equation}
For the stationary case, the phase function $S$ can be written as
$S=-\omega t+W(r,\theta)$,
\begin{align}
  &\tilde k^i=(k^r,k^\theta,0,\cdots)=\left(f
    W_r,\frac{1}{r^2}W_\theta,0,\cdots\right),\quad \tilde k^i\tilde
    k_i=\frac{\omega^2}{f},\\
  & f^{-1/2}D_i\left(a^2 f^{1/2}\tilde k^i\right)=0. \label{eq:KGJ2}
\end{align}
Here, $\tilde k^i$ represents the tangent vector of null
rays projected on a constant time slice. We can write the solution
of \eqref{eq:KGJ} as
\begin{equation}
  a(\lambda,\chi)=a(\lambda_0,\chi)\exp\left(-\frac{1}{2}\int_{\lambda_0}^\lambda
    d\lambda\,\Theta(\lambda)\right),\quad\Theta=\nabla_\mu k^\mu, 
\end{equation}
where the integral is along a null ray (with respect to the affine
parameter $\lambda$) and $\chi$ denotes a coordinate distinguishing
different geodesics. As the expansion of null congruence from the AdS
boundary is zero, the amplitude $a(\lambda,\chi)$ is conserved along a
null ray and independent of $\lambda$. Furthermore, for a point source
isotropically emitting null rays, $a$ is independent of $\chi$ and can
assume to be constant. Thus \eqref{eq:KGJ2} implies
  \begin{equation}
    D_i\tilde n^i=0,\quad \tilde n^i=\frac{f^{1/2}}{\omega}\tilde k^i,
    \quad \tilde
    n^i\tilde n_i=1,
  \end{equation}
  and $\tilde n^i$ is divergenceless normalized vector field.  The
  wave front is the surface with the unit normal $\tilde n^i$, and is
  the extremal surface. The number of null rays passing through the
  wave front $\mathcal{E}_A$, which is the extremal surface homologous to the
  region $A$ on the AdS boundary, is
  \begin{equation}
    \mathrm{Area}(\mathcal{E}_A)=\int_{\mathcal{E}_A}\tilde n^i d\Sigma_i,\quad d\Sigma_i=\tilde
    n_i\sqrt{h}\,d^{d-1}\sigma, \label{eq:RTflux}
  \end{equation}
where $\sqrt{h}$ denotes determinant of the induced metric on
$\mathcal{E}_A$. Now let us consider the setup shown as
Fig.~\ref{fig:screen-setup}. We prepare a screen $A(\ep)$ which is
$r=$constant surface in the bulk. For the regularization, the screen is
placed at $r=\lad^2/\ep$ near the AdS boundary.
\noindent
Because $\tilde n^i$ is divergence free vector field,
Eq.~\eqref{eq:RTflux} equals to
\begin{equation}
  \mathrm{Area}(\mathcal{E}_A)=\int_{A(\ep)}\tilde n^i
  d\Sigma_i=\frac{1}{\omega}\int_{A(\ep)}J^id\Sigma_i.
   \label{eq:flux-formula}
\end{equation}
This is a formula for area of the RT surface in terms of flux
integration of null rays on the screen $A(\ep)$. As the
  Klein-Gordon current $J^i/\omega=f^{1/2}\tilde k^i/\omega$
  represents the number density of null rays, we can regard the
  Klein-Gordon current as  a representation of the amount of
  information propagating in the bluk from the AdS bounary. 

As a demonstration, we evaluate the right hand side of this relation
for the BTZ spacetime.  By fixing the radial coordinate as
$r =\lad^2 / \epsilon $ in Eq.~\eqref{Null geodesic Planar BTZ in
  Examples}, the impact parameter $b$ on the screen is
\begin{align}
b=\frac{ \lad^2 \sinh \left( \sqrt{M}\, \theta\right)}{\sqrt{\epsilon^2 M
  + \lad^2 \sinh^2 \left( \sqrt{M}\, \theta\right)}}. \label{eq:impact}
\end{align}
From   Eq.~\eqref{eq:geod1}, the radial component of the tangent vector of
the null ray is
\begin{align}
\frac{\tilde k^r}{\omega} =\left(1-f\,\frac{b^2}{r^2}\right)^{1/2}
= \sqrt{1- \left(\frac{r^2}{\lad^2}-M\right)\frac{b^2}{r^2}},
\label{kr on screen in Geodesical viewpoint}
\end{align}
and on the screen,
\begin{equation}
  \left.\frac{\tilde k^r}{\omega}\right|_{A(\epsilon)}=\frac{ \sqrt{M} \cosh \left( \sqrt{M}\, \theta\right)}{\sqrt{ M + (\lad^2/\ep^2) \sinh^2 \left( \sqrt{M}\, \theta\right)}}.
\end{equation}
The area element on the screen is
\begin{align}
\left.  d \Sigma_r\right|_{A(\epsilon)}
= \left.r\,n_r\,d\theta\right|_{A(\epsilon)}
= r f^{-1/2}d\theta,
\end{align}
where $n_r$ is the radial component of the unit normal to the screen. Thus
\begin{equation}
 f^{1/2}\frac{\tilde k^r}{\omega} \left.d\Sigma_r\right|_{A(\ep)}
 =r\left(1-f\, \frac{b^2}{r^2}\right)^{1/2}\!\!\!\!d\theta
 =\frac{\sqrt{M} \lad \cosh(\sqrt{M}\theta)d\theta}{\sqrt{\sinh^2(\sqrt{M}\theta)+M\ep^2/\lad^2}}.
\end{equation}
Therefore, \eqref{eq:flux-formula} become 
\begin{align}
\int_{-\theta_\ell}^{\theta_\ell} \frac{f^{1/2}\tilde k^i d\Sigma_i }{\omega} 
&= \int_{-\theta_\ell}^{\theta_\ell} d\theta \frac{\sqrt{M} \lad \cosh \left(
  \sqrt{M}\, \theta \right)}{\sqrt{M\epsilon^2/\lad^2 +  \sinh^2
  \left( \sqrt{M}\, \theta \right)}} \nonumber\\ 
& = \lad \log \left[ 
 \frac{ \sqrt{\sinh^2(\sqrt{M}\, \theta_\ell)+M\ep^2/\lad^2}+\sinh(\sqrt{M}\theta_\ell)}{\sqrt{\sinh^2(\sqrt{M}\, \theta_\ell)+M\ep^2/\lad^2}-\sinh(\sqrt{M}\theta_\ell)}
  \right] \nonumber\\
& = 2 \lad \log\left[ \frac{\lad}{(\epsilon/2) \sqrt{M}} \sinh \left(
  \sqrt{M}\, \theta_\ell\right) \right]+O(\epsilon),
\end{align}
and reproduces the ``area'' of the RT surface~\eqref{eq:finite-temp}. Dividing by $4G_N$,
 this result correctly reproduces the EE of
CFT \eqref{eq:EET}.  Therefore,
we can regard such a null geodesic congruence as one realization of
the bit threads.


\section{Summary}

In this paper, we show that wave fronts of null rays emitted
  from a point on the AdS boundary are extremal surfaces  in static
  spherical symmetric spacetimes. Thus the RT surface can be
  understood as a wave front, and null rays naturally define a flow
   characterizing the amount of  the EE of CFT. Hence such a flow
  can be regarded as the bit threads. 

  As we assumed a point source on the AdS boundary, the shape of a
  region on the AdS boundary (entangling surface) becomes spherical
  because the boundary of the region is a wave front on the AdS
  boundary. However, by superposing point sources, it is possible to
  construct an extremal surface homologous to a region with arbitrary
  shapes on the AdS boundary by considering the envelope of wave
  fronts from each point sources. Thus the method presented in this
  paper may be applicable to the plateaux problem~\cite{Hubeny2013,
    Freivogel2015} with non-trivial shapes of an entangling surface
  and to further understanding of property of the holographic EE.

\acknowledgments{
Y. N. was supported in part by JSPS KAKENHI Grant Number
19K03866.}

\appendix
\section{Massless scalar field  in AdS spacetimes}
We consider the solution of the Klein-Gordon equation
$\square\phi=0$ in the AdS spacetime with the metric
\eqref{eq:metric-Ddim}. Assuming  the axially symmetric and stationary configuration of the scalar field $\phi=e^{-i\omega t}\tilde\phi(r,\theta)$, $\tilde\phi$ obeys the following Helmholtz type equation
\begin{equation}
\frac{\pa^2\tilde\phi}{\pa r^2}+\left(\frac{d-1}{r}+\frac{f'}{f}\right)\frac{\pa\tilde\phi}{\pa r}+\frac{\omega^2}{f^2}\tilde\phi+
\frac{1}{fr^2\sin^{d-2}\theta}\frac{\pa}{\pa\theta}\left(\sin^{d-2}\theta\,\frac{\pa\tilde\phi}{\pa\theta}\right)=0.
\end{equation}
Assuming $\tilde\phi=R(r)\Phi(\theta)$,
\begin{align}
&R''+\left(\frac{d-1}{r}+\frac{f'}{f}\right)R'+\left(\frac{\omega^2}{f^2}
-\frac{m(m+d-2)}{fr^2}\right)R=0, \\
&\Phi_{\theta\theta}+(d-2)\cot\theta\,\Phi_\theta+m(m+d-2) \Phi=0,
\end{align}
and $\Phi=C_m^{d/2-1}(\cos\theta)$ (Gegenbauer polymonial).  For
$d=2$, $\Phi=e^{im\theta},~m\in \mathbb{Z}$ and for $d=3$,
$\Phi=P_m(\cos\theta),~m \in \mathbb{Z}_+$.  We consider $d=2$
case. For the normalized radial function
$\lim_{r\rightarrow\infty}R_m(r)=1$, the solution of the Klein-Gordon
equation with a point source at the AdS boundary is represented as
\begin{equation}
  \phi(r,\theta)=\sum_{m=-\infty}^{m=\infty}e^{im\theta}R_m(r), \label{eq:app-eq1}
\end{equation}
and this wave function gives
$\lim_{r\rightarrow\infty}\phi\propto\del(\theta)$ and satisfies the
boundary condition with a point wave source at the AdS boundary.  For
the BTZ spacetime, the solution satisfying the ingoing boundary
condition at the black hole horizon is given by
\begin{align}
  &R_m=\frac{\Gamma(a)\Gamma(b)}{\Gamma(c)}\xi^{-i\frac{\lad\omega}{2\sqrt{M}}}F\left[
    \frac{i}{2\sqrt{M}}(\lad\,\omega-m), \frac{i}{2\sqrt{M}}(\lad\,\omega+m), 
    1+i\frac{\lad\,\omega}{\sqrt{M}},\xi\right],\\
  &\quad a=1-\frac{i}{2\sqrt{M}}(\lad\,\omega-m),\quad
    b=1+\frac{i}{2\sqrt{M}}(\lad\,\omega-m),
    \quad c=1+i\frac{\lad\,\omega}{\sqrt{M}}, \notag
\end{align}
where $F$ is Gauss's hypergeometric function and
$\xi=1-M\lad^2/r^2$. The figures~\ref{fig:wave-ads} and
\ref{fig:wave-BTZ} are obtained by taking sum in \eqref{eq:app-eq1} up
to $m_\text{max}\sim 200$.


\end{document}